**Analyzing PFG anisotropic anomalous diffusions by instantaneous signal attenuation method**

Guoxing Lin*

Carlson School of Chemistry and Biochemistry, Clark University, Worcester, MA 01610, USA

**Abstract**

Anomalous diffusion has been investigated in many systems. Pulsed field gradient (PFG) anomalous diffusion is much more complicated than PFG normal diffusion. There have been many theoretical and experimental studies for PFG isotropic anomalous diffusion, but there are very few theoretical treatments reported for anisotropic anomalous diffusion. Currently, there is not a general PFG signal attenuation expression, which includes the finite gradient pulse effect and can treat all three types of anisotropic fractional diffusions, which can be classified by time derivative order $\alpha$ and space derivative order $\beta$ as: general fractional diffusion $\{0 < \alpha, \beta \leq 2\}$, time fractional diffusion $\{0 < \alpha \leq 2, \beta = 2\}$, and space-fractional diffusion $\{\alpha = 1, 0 < \beta \leq 2\}$. In this paper, the recently developed instantaneous signal attenuation (ISA) method was applied to obtain PFG signal attenuation expression for free and restricted anisotropic anomalous diffusion with two models: fractal derivative and fractional derivative models. The obtained PFG signal attenuation expression for anisotropic anomalous diffusion can reduce to the reported result for PFG anisotropic normal diffusion when $\alpha = 1$ and $\beta = 2$. The results can also reduce to reported PFG isotropic anomalous diffusion results obtained by effective phase shift diffusion equation method and instantaneous signal attenuation method. For anisotropic space-fractional diffusion, the obtained result agrees with that obtained by the modified Bloch equation method. Additionally, The PFG signal attenuation expressions for free and restricted anisotropic curvilinear diffusions were derived by the traditional method, the results of which agree with the PFG anisotropic fractional diffusion results based on the fractional derivative model. The powder pattern of PFG anisotropic diffusion was also discussed. The results here improve our understanding of PFG anomalous diffusion, and provide new formalisms for PFG anisotropic anomalous diffusion in NMR and MRI.

**1. Introduction**

Anomalous diffusion [1,2,3] can be found in many systems such as polymer or biological systems [4], porous materials [5,6], fractal geometries [7], micelle systems [8] and single-file structures [9,10]. Some anomalous diffusions could be modeled by time-space fractional diffusion equations with time derivative order $\alpha$ and space derivative order $\beta$, $0 < \alpha, \beta \leq 2$ [1, 2, 11, 12, 13, 14, 15, 16]; $\alpha > 1$ corresponds to superdiffusion, while $\alpha < 1$ is referred to as subdiffusion [1, 2]. There are three types of fractional diffusions: general fractional diffusion $\{0 < \alpha, \beta \leq 2\}$, space-fractional diffusion $\{\alpha = 1, 0 < \beta \leq 2\}$, and time-fractional diffusion $\{0 < \alpha \leq 2, \beta = 2\}$ [11-16]. Pulsed field gradient (PFG) [17,18,19,20,21] technique is important tool to monitor normal and anomalous diffusion. There have been many experimental and theoretical studies of anomalous diffusion. Some of these studies include Kärger et al.'s Gaussian phase distribution (GPD) approximation with autocorrelation function [22,23], the propagator representation [24], the short gradient pulse (SGP) approximation [25], the stretched exponential models [26, 27], the modified Bloch equations [28,29,30], the log-normal distribution function [31], the recently developed effective phase shift diffusion equation (EPSDE) method [16], the instantaneous signal attenuation (ISA) method [32], the modified GPD approximation method [33], and the non-GPD approximation method [34]. These studies have yielded encouraging results for PFG isotropic anomalous diffusion in both free and restricted diffusion. Currently, there are the expressions for all three types of free isotropic diffusions [16,22,28,32,33,34]. Meanwhile, the PFG restricted anomalous diffusions inside plate, sphere and cylinder have been investigated [35]; an apparent anomalous diffusion coefficient $D_{f,app}(t) = D_{f_0}\left[1 - c_f \frac{S}{V}\left(D_{f_0} t^\alpha\right)^{1/\beta}\right]$ has also been proposed for restricted diffusion [35]. Additionally, it is found that the conventional PFG methods such as the time-dependent diffusivities may not be sufficient in analyzing PFG anomalous diffusion, and the parameters associated with the degree of non-Gaussian behavior are more sensitive to ischemic changes than the apparent normal diffusion coefficient [36]. Compared to the large amount of PFG isotropic diffusion studies, the progress in the research of anisotropic anomalous diffusion is slow.

There are very limited studies on PFG anisotropic anomalous diffusion. In real materials, the anomalous diffusion could be anisotropic rather than isotropic. The cause of anisotropic behavior for anomalous diffusion could be similar to that of anisotropic normal diffusion, which results from the physical arrangement or anisotropic restriction [21]. The anisotropic diffusion could be observed in many materials [37, 38, 39] such as stretched or compressed polymers, block polymers, liquid crystals, muscle cells, brain tissues and so on. Some of these anisotropic diffusions could be anomalous diffusions. The conventional anisotropic normal diffusion formalisms may not be sufficient to analyze these anisotropic anomalous diffusions. There are very few reported theoretical studies about anisotropic anomalous diffusion. These theoretical PFG anisotropic anomalous diffusion studies have been investigated by the modified Bloch equation methods [29,40,41] Thus, it is necessary to develop some new treatments for PFG anisotropic anomalous diffusion.

In this paper, to address these challenges for PFG anisotropic anomalous diffusion, the recently developed instantaneous signal attenuation method [32] for isotropic anomalous diffusion was extended straightforwardly to analyze the PFG anisotropic fractional diffusion. The ISA method is an intuitive method, and it has been applied to obtain two general PFG signal attenuation expressions for free isotropic anomalous diffusion: a stretched exponential attenuation $\exp\left[-D_{f_1}\int_0^t K^\beta(t')dt'^\alpha\right]$ from the fractal derivative (defined in Appendix A.) model [16, 32] and a Mittag-Leffler function attenuation $E_\alpha\left[-D_{f_2}\int_0^t K^\beta(t')dt'^\alpha\right]$ [16,32] from the fractional derivative (defined in Appendix B.) model, where $E_{\alpha,1}(x) = \sum_{n=0}^\infty \frac{x^n}{\Gamma(n\alpha+1)}$ is the Mittag-Leffler function, $D_{f_1}$ and $D_{f_2}$ are the fractional diffusion coefficients with units $m^\beta/s^\alpha$ for fractal derivative and fractional derivative respectively, and $K(t') = \int_0^{t'} \gamma g(t'')dt''$ is the wavenumber [20,21]. The instantaneous signal attenuation method has also been applied to study PFG restricted isotropic free anomalous diffusion as well as restricted anomalous diffusion [32].

In this paper, the instantaneous signal attenuation method was used to obtain theoretical signal attenuation expressions for free or restricted anisotropic fractional diffusions. Two models including fractal derivative model [11,12] and fractional derivative model [13-15] were used. The results from these two

Guoxing Lin, PFG signal attenuation of anisotropic anomalous diffusion    2

models are complementary to each other [16,32,35]. At small signal attenuation, the results from these two models are approximately equal for all three types of anisotropic diffusions. For space-fractional diffusion, the results from the two models are the same and agree with that reported by the modified fractional Bloch equation method [28]. The obtained anisotropic anomalous diffusion results can reduce to not only the reported isotropic anomalous diffusion result obtained from ISA and EPSDE methods [16,32], but also the reported anisotropic normal diffusion result [42]. Additionally, the free and restricted anisotropic curvilinear diffusions were analyzed by traditional methods, whose results agree with the results of PFG anisotropic anomalous diffusion based on the fractional derivative model. The results here help us understand PFG anisotropic anomalous diffusion.

**2. Theory**

2.1. Brief introduction of ISA method [32]

The recently developed ISA method has been applied to describe PFG isotropic anomalous diffusion [32]. The ISA method is modified from the propagator approach for normal diffusion [43]. The fundamental idea of ISA approximation method is intuitive. In a macroscopic sample with an enormous number of spin carriers, the phase difference of the spin carriers diffusing to the same location could be averaged out and could not be distinguished after mixing for every moment, which results in an instantaneous signal attenuation; the total signal attenuation is the cumulative results of the whole diffusion processes [32,43].

The ISA can be obtained by the following process [16]: first, obtain the signal attenuation by the short gradient pulse approximation; second, get the instantaneous signal attenuation $a(K_{SGP}, t', dt')$ by

$$a(K_{SGP}, t', dt') = A_{SGP}(K_{SGP}, t' + dt') / A_{SGP}(K_{SGP}, t'), \qquad (1)$$

where $A_{SGP}$ is the signal attenuation under SGP approximation, and $K_{SGP}$ is the wavenumber; third, by replacing $K_{SGP}$ with time-dependent $K(t')$, the instantaneous signal attenuation $a(K(t'), t', dt')$ can be obtained as

$$a(K_{SGP}, t', dt') \xrightarrow{K_{SGP} \leftrightarrow K(t')} a(K(t'), t', dt'), \qquad (2a)$$

$$a(K(t'), t', dt') = \begin{cases} A(K(t'), t' + dt') / A(K(t'), t') \\ 1 + \dfrac{\left[\dfrac{\partial}{\partial t'} A(K(t'), t')\right] dt'}{A(K(t'), t')} = 1 + \ln[A(K(t'), t' + dt')] - \ln[A(K(t'), t')] \end{cases}. \qquad (2b)$$

In Eq. (2b), $K(t')$ is treated as a 'constant' during the short time interval $dt'$, which is reasonable because the instantaneous signal attenuation is caused by spatially averaging resulting from the diffusion process. The reason for this treatment is clear from the EPSDE method [16]. In the EPSDE method, at the time $t'$, the accumulating phase shift change depends on $K(t')\Delta(z)$, where $\Delta(z)$—the small random displacement due to diffusion—is important and $\Delta K(t')\Delta(z)$ is a neglectable second order change during the infinitesimal time interval, $dt'$ [32].

The net signal attenuation including finite gradient width effect will be the cumulative product of all instantaneous signal attenuations,

$$A(t) = \prod_{0 < t' < t, dt'} a(K(t'), t', dt'). \qquad (3)$$

This method has been used to obtain the theoretical signal attenuation expressions for free fractional diffusion, restricted fractional diffusion and fractional diffusion with nonlinear field [32].

2.2. Free fractional diffusion

2.2.1 General anisotropic fractional diffusion

Here the ISA method [32] is used to analyze PFG anisotropic fractional diffusion with two models: the fractal derivative [11,12] and the fractional derivative [13-15] models.

2.2.1.1 Fractal derivative model

The fractal derivative model has been developed in 2006 by Chen [11,12]. In principal frame, the three-dimensional anisotropic anomalous diffusion may be written as

$$\frac{\partial P'(x',y',z',t)}{\partial t^\alpha} = D'_{f_1 x'x'} \frac{\partial}{\partial x'^{\beta_{x'}/2}} \left( \frac{\partial P'(x',y',z',t)}{\partial x'^{\beta_{x'}/2}} \right) +$$
$$D'_{f_1 y'y'} \frac{\partial}{\partial y'^{\beta_{y'}/2}} \left( \frac{\partial P'(x',y',z',t)}{\partial y'^{\beta_{y'}/2}} \right) + D'_{f_1 z'z'} \frac{\partial}{\partial z'^{\beta_{z'}/2}} \left( \frac{\partial P'(x',y',z',t)}{\partial z'^{\beta_{z'}/2}} \right).$$

(4)

where the fractal derivative has been used (see Appendix A and references [11,12]), $D'_{f_1 ii}$, $i = x', y', z'$ is the general fractional diffusion coefficient with units of $m^{\beta_i}/s^\alpha$, and $\beta_i$ is the fractal derivative order. When $D'_{f_1 x'x'} = D'_{f_1 y'y'} = D'_{f_1 z'z'} = D_{f_1}$ and $\beta_{x'} = \beta_{y'} = \beta_{z'} = \beta$, the anisotropic diffusion reduces to isotropic diffusion. The strategy to solve Eq. (4) is like that used to solve one dimensional fractional diffusion equation in references [11]. Based on Refs. [11], by introducing

$$\begin{cases} \theta = t^\alpha \\ \xi_i = x'^{\beta_{x'}/2} \\ \xi_j = y'^{\beta_{y'}/2} \\ \xi_k = z'^{\beta_{z'}/2} \end{cases}, \quad (5)$$

the following equation is obtained:

$$\frac{\partial P'}{\partial \theta} = D'_{f_1 x'x'} \frac{\partial}{\partial \xi_i}\left(\frac{\partial P'}{\partial \xi_i}\right) + D'_{f_1 y'y'} \frac{\partial}{\partial \xi_j}\left(\frac{\partial P'}{\partial \xi_j}\right) + D'_{f_1 z'z'} \frac{\partial}{\partial \xi_k}\left(\frac{\partial P'}{\partial \xi_k}\right). \quad (6)$$

The solution of Eq. (6) is [44]

$$P' = \frac{1}{(4\pi\theta)^{3/2} \sqrt{D'_{f_1 x'x'} D'_{f_1 y'y'} D'_{f_1 z'z'}}} \exp\left[-\frac{\xi_i^2}{4D'_{f_1 x'x'}\theta} - \frac{\xi_j^2}{4D'_{f_1 y'y'}\theta} - \frac{\xi_k^2}{4D'_{f_1 z'z'}\theta}\right]. \quad (7)$$

where the initial position of the self-diffusion is set at the origin of the coordinate system. Substituting Eq. (5) into Eq. (7), after normalization, we get

$$P' = \frac{c}{(4\pi t^\alpha)^{3/2}\sqrt{D'_{f_1 x'x'} D'_{f_1 y'y'} D'_{f_1 z'z'}}} \exp\left[-\frac{x'^{\beta_{x'}}}{4D'_{f_1 x'x'}t^\alpha} - \frac{y'^{\beta_{y'}}}{4D'_{f_1 y'y'}t^\alpha} - \frac{z'^{\beta_{z'}}}{4D'_{f_1 z'z'}t^\alpha}\right], \quad (8)$$

where $c$ is a normalization factor. In Chen's paper [11,12], the Fourier transform of the solution of one dimensional diffusion is $\int_{-\infty}^{\infty} P(z,t)\exp(+ikz)dz = \exp(-D_{f_1} k^{\beta_{z'}} t^\alpha)$, based on which the signal attenuation under SGP approximation can be obtained as

$$A_{SGP}(t) = \int_{-\infty}^{\infty} P' \exp(-i\gamma\delta \mathbf{g}' \cdot \mathbf{r}') dV'$$

$$= \exp\left[-\sum_{i=x',y',z'} g_i'^{\beta_i/2} D'_{f_1 ii} g_i'^{\beta_i/2} (\gamma\delta)^{\beta_i/2} t^\alpha\right]$$

$$= \exp\left[-\sum_{i=x',y',z'} D'_{f_1 ii} (\gamma g_i' \delta)^{\beta_i} t^\alpha\right], \quad (9)$$

$$= \exp\left(-\sum_{i=x',y',z'} D'_{f_1 ii} K'^{\beta_i}_{SGP} t^\alpha\right)$$

where $K'_{SGP} = \gamma g'_i \delta$ is the wavenumber, and $\mathbf{g}'$ is the effective gradient in the principal-axis frame, which can be obtained by

$$\mathbf{g}' = R\mathbf{g}, \quad (10)$$

where $\mathbf{g}$ is the gradient vector in the observational reference frame, and $R$ is the rotation matrix that rotates vectors or tensors from the observational reference frame to the principal frame. Based on Eqs. (1)-(3), the signal attenuation including finite gradient width effect can be calculated as

$$A(t) = \exp\left(-\sum_{i=x',y',z'} \int_0^t D'_{f_1 ii} K'^{\beta_i}_i(t') dt'^\alpha\right), \quad (11)$$

where $K'_i(t')$ is the wavenumber in the principal frame, which is defined as

$$K'_i(t') = \int_0^{t'} \gamma g'_i(t'') dt''. \quad (12)$$

When $D'_{f_1 x'x'} = D'_{f_1 y'y'} = D'_{f_1 z'z'} = D_{f_1}$ and $\beta_{x'} = \beta_{y'} = \beta_{z'} = \beta$, Eq. (11) reduces to the signal attenuation for an isotropic fractional diffusion

$$A(t) = \exp\left\{-D_{f_1}\left[\sum_{i=x,y,z} \int_0^t K_i^\beta(t') dt'^\alpha\right]\right\}. \quad (13)$$

If the gradient is only applied in one dimension, Eq. (13) reduces to

$$A(t) = \exp\left[-D_{f_1} \int_0^t K^\beta(t') dt'^\alpha\right], \quad (14)$$

which is the same as that obtained from the effective phase shift diffusion method [16] as well as the ISA method [32].

Eq. (11) can be further written as

$$A(t)_{free} = \exp(-\mathbf{b}'* : \mathbf{D}'_{\mathbf{f_1}}) = \exp(-\int_0^t \widetilde{\mathbf{K}}^*(\mathbf{t}') \cdot \mathbf{D}'_{\mathbf{f_1}} \cdot \mathbf{K}^*(\mathbf{t}') dt'^\alpha), \quad (15)$$

where $\mathbf{K}^*(\mathbf{t}')$ is

$$\mathbf{K}^*(\mathbf{t}') = \begin{pmatrix} K'^{\beta_{x'}/2}_{x'}(t)\mathbf{i}_{x'} \\ K'^{\beta_{y'}/2}_{y'}(t)\mathbf{i}_{y'} \\ K'^{\beta_{z'}/2}_{z'}(t)\mathbf{i}_{z'} \end{pmatrix} = \begin{pmatrix} \left[\sum_{j=x,y,z} R_{xj} K_x(t)\right]^{\beta_{x'}/2} \mathbf{i}_{x'} \\ \left[\sum_{j=x,y,z} R_{yj} K_y(t)\right]^{\beta_{y'}/2} \mathbf{i}_{y'} \\ \left[\sum_{j=x,y,z} R_{zj} K_z(t)\right]^{\beta_{z'}/2} \mathbf{i}_{z'} \end{pmatrix},$$

(16)

and

$$\mathbf{b}'^* = \int_0^t \widetilde{\mathbf{K}}^*(\mathbf{t}') \otimes \mathbf{K}^*(\mathbf{t}') dt'^\alpha \ . \tag{17}$$

The attenuation can be rewritten as

$$\begin{aligned} A(t) &= \exp(-\int_0^t \widetilde{\mathbf{K}}^*(\mathbf{t}') \cdot \mathbf{D}'_{\mathbf{f}_1} \cdot \mathbf{K}^*(\mathbf{t}') dt'^\alpha) \\ &= \exp(-\int_0^t \widetilde{\mathbf{K}}^*(\mathbf{t}') \cdot R \cdot \mathbf{D}_{\mathbf{f}_1} \cdot \widetilde{R} \cdot \mathbf{K}^*(\mathbf{t}') dt'^\alpha), \\ &= \exp(-\mathbf{b}^* : \mathbf{D}_{\mathbf{f}_1}) \end{aligned} \tag{18}$$

where $\mathbf{D}_{\mathbf{f}_1} = \widetilde{R} \cdot \mathbf{D}'_{\mathbf{f}_1} \cdot R$, and

$$\mathbf{b}^* = \int_0^t \left[\widetilde{\mathbf{K}}^*(\mathbf{t}') \cdot R\right] \otimes \left[\widetilde{R} \cdot \mathbf{K}^*(\mathbf{t}')\right] dt'^\alpha \ . \tag{19}$$

For normal anisotropic diffusion, $\alpha = 1, \beta = 2$, $\mathbf{K}^*(\mathbf{t}') = R\mathbf{K}(\mathbf{t}') = \mathbf{K}'(\mathbf{t}')$, $\mathbf{b}'^* = \mathbf{b}'$, $\mathbf{b}^* = \mathbf{b}$, and Eqs. (17) and (18) reduce to normal anisotropic diffusion result [21,42]

$$A(t) = \exp(-\mathbf{b}' : \mathbf{D}') = \exp(-\mathbf{b} : \mathbf{D}) \ . \tag{20}$$

2.2.1.2 Fractional derivative model

The multiple-dimensional fractional diffusion has been studied by fractional derivative model in many references [29,40,41,45,46]. Based on Ref. [29,41,45], the anisotropic fractional diffusion equation may be written as

$$_t\mathbf{D}^\alpha_* P'(x',y',z',t) = \left( D'_{f_2 x' x'} \frac{\partial^{\beta_{x'}}}{\partial |x'|^{\beta_{x'}}} + D'_{f_2 y' y'} \frac{\partial^{\beta_{y'}}}{\partial |y'|^{\beta_{y'}}} + D'_{f_2 z' z'} \frac{\partial^{\beta_{z'}}}{\partial |z'|^{\beta_{z'}}} \right) P'(x',y',z',t), \tag{21}$$

where $\frac{\partial^{\beta_{x'}}}{\partial |x'|^{\beta_{x'}}}$, $\frac{\partial^{\beta_{y'}}}{\partial |y'|^{\beta_{y'}}}$ and $\frac{\partial^{\beta_{z'}}}{\partial |z'|^{\beta_{z'}}}$ are fractional derivatives defined in Appendix B. [3,35,45]. Applying Fourier transform to Eq. (21) gives

$$_t\mathbf{D}_*^\alpha P'(k'_{x'},k'_{y'},k'_{z'},t) = -\left(D'_{f_2 x'x'}|k'_{x'}|^{\beta_{x'}} + D'_{f_2 y'y'}|k'_{y'}|^{\beta_{y'}} + D'_{f_2 z'z'}|k'_{z'}|^{\beta_{z'}}\right)P'(k'_{x'},k'_{y'},k'_{z'},t), \quad (22)$$

where $P'(k'_{x'},k'_{y'},k'_{z'},t)$ is the probability density function in the $k'_{x'}, k'_{y'}, k'_{z'}$ space. The solution of Eq. (22) is [1,2,13,15]

$$P'(k'_{x'},k'_{y'},k'_{z'},t) = E_{\alpha,1}\left[-\left(D'_{f_2 x'x'}|k'_{x'}|^{\beta_{x'}} + D'_{f_2 y'y'}|k'_{y'}|^{\beta_{y'}} + D'_{f_2 z'z'}|k'_{z'}|^{\beta_{z'}}\right)t^\alpha\right]. \quad (23)$$

The signal attenuation can be calculated as [16,20,21]

$$A_{f_2,SGP}(t) = \int_{-\infty}^{+\infty} P(x',y',z',t)\exp(\mathbf{K}'_{SGP} \cdot \mathbf{r}')dV' = P'(K'_{x'},K'_{y'},K'_{z'},t) = E_{\alpha,1}\left(-\sum_{i=x',y',z'} D'_{f_2 ii} K'^{\beta_i}_{SGP,i} t^\alpha\right). \quad (24)$$

Based on Eqs. (1)-(3), the signal attenuation including the finite gradient pulse width effect will be

$$A(t) = E_{\alpha,1}\left(-\sum_{i=x',y',z'} \int_0^t D'_{f_2 ii} K'^{\beta_i}_i(t')dt'^\alpha\right). \quad (25)$$

Similarly, Eq. (25) can be reduced to give a signal attenuation expression for isotropic three-dimensional diffusion, which is

$$A(t) = E_{\alpha,1}\left\{-D_{f_2}\left[\sum_{i=x,y,z}\int_0^t K_i^\beta(t')dt'^\alpha\right]\right\}. \quad (26)$$

If the gradient is only applied in one dimension, Eq. (26) reduces to

$$A(t) = E_{\alpha,1}\left[-D_{f_2}\int_0^t K^\beta(t')dt'^\alpha\right], \quad (27)$$

which reproduces the ISA method result in Ref. [32].

Comparing Eq. (26) with Eq. (11), we can see that the difference between the results from the fractional derivative and the fractal derivative is the attenuation function. One is a stretched exponential function. The other is a Mittag-Leffler function. The two terms $-\sum_{i=x',y',z'}\int_0^t D'_{f_1 ii}K'^{\beta_i}_i(t')dt'^\alpha$ and $-\sum_{i=x',y',z'}\int_0^t D'_{f_2 ii}K'^{\beta_i}_i(t')dt'^\alpha$ are similar for these two attenuation expressions. Eq. (26) can be further written in diffusion tensor form as

$$A(t) = E_{\alpha,1}(-\int_0^t \tilde{\mathbf{K}}^*(\mathbf{t'}) \cdot \mathbf{D}'_{\mathbf{f_2}} \cdot \mathbf{K}^*(\mathbf{t'})dt'^\alpha) = E_{\alpha,1}(-\mathbf{b}'* : \mathbf{D}'_{\mathbf{f_2}})$$
$$= E_{\alpha,1}(-\int_0^t \tilde{\mathbf{K}}^*(\mathbf{t'}) \cdot R \cdot \mathbf{D}_{\mathbf{f_2}} \cdot \tilde{R} \cdot \mathbf{K}^*(\mathbf{t'})dt'^\alpha) \quad , \quad (28)$$
$$= E_{\alpha,1}(-\mathbf{b}* : \mathbf{D}_{\mathbf{f_2}})$$

where $\mathbf{K}^*(\mathbf{t'})$, $\mathbf{b}'$ and $\mathbf{b}'*$ are the same as these defined in subsection 2.2.1.1 for fractal derivative model. When $-\mathbf{b}'* : \mathbf{D}'_{\mathbf{f_2}}$ is small, Eq. (28) reduces to

$$A(t) = E_{\alpha,1}(-\mathbf{b}'^* : \mathbf{D}_{f_2}) \approx \exp\left[-\mathbf{b}'^* : \mathbf{D}'_{f_2} / \Gamma(1+\alpha)\right]. \tag{29}$$

which reproduces fractal derivative result Eq. (15) if we could set

$$\mathbf{D}'_{f_1} = \mathbf{D}'_{f_2} / \Gamma(1+\alpha). \tag{30}$$

2.2.2 Anisotropic time-fractional diffusion

For a time-only anisotropic fractional diffusion, it has $0 < \alpha \leq 2$, and $\beta_i = 2$, $i = x', y', z'$, from Eqs. (11) and (25), the PFG signal attenuation will be

$$A(t) = \begin{cases} \exp\left\{-\left[\sum_{i=x',y',z'} D'_{f_1 ii} \int_0^t k_i'^2(t') dt'^\alpha\right]\right\} = \exp(-\mathbf{b}'_t{}^* : \mathbf{D}'_{f_1}) = \exp(-\mathbf{b}_t{}^* : \mathbf{D}_{f_1}), & \text{fractal derivative} \\ E_{\alpha,1}\left\{-\left[\sum_{i=x',y',z'} D'_{f_2 ii} \int_0^t k_i'^2(t') dt'^\alpha\right]\right\} = E_{\alpha,1}(-\mathbf{b}'_t{}^* : \mathbf{D}'_{f_2}) = E_{\alpha,1}(-\mathbf{b}_t{}^* : \mathbf{D}_{f_2}), & \text{fractional derivative} \end{cases}, \tag{31}$$

where

$$\begin{aligned}
\mathbf{b}_t^* &= \int_0^t \left[\tilde{\mathbf{K}}^*(\mathbf{t}') \cdot R\right] \otimes \left[\tilde{R} \cdot \mathbf{K}^*(\mathbf{t}')\right] dt'^\alpha \\
&= \int_0^t \left[\tilde{\mathbf{K}}'(\mathbf{t}') \cdot R\right] \otimes \left[\tilde{R} \cdot \mathbf{K}'(\mathbf{t}')\right] dt'^\alpha \\
&= \int_0^t \left[\tilde{\mathbf{K}}(\mathbf{t}')\tilde{R} \cdot R\right] \otimes \left[\tilde{R} \cdot R\mathbf{K}(\mathbf{t}')\right] dt'^\alpha \\
&= \int_0^t \tilde{\mathbf{K}}(\mathbf{t}') \otimes \mathbf{K}(\mathbf{t}') dt'^\alpha
\end{aligned} \tag{32}$$

where $\tilde{\mathbf{K}}(\mathbf{t}') \otimes \mathbf{K}(\mathbf{t}')$ is the same as that of anisotropic normal diffusion, which is reasonable as both time-fractional diffusion and normal diffusion have $\beta = 2$.

2.2.3 Anisotropic space-fractional diffusion

For a space-fractional diffusion, $\alpha = 1$, and $0 < \beta_i \leq 2, i = x', y', z'$. When $\alpha = 1$, $E_{1,1}(x) = \exp(x)$ therefore, the signal attenuations from fractal derivative and fractional derivative are the same, which is

$$A(t) = \exp\left\{-\left[\sum_{i=x',y',z'} D'_{f_1 \text{or} 2 ii} \int_0^t K_i'^\beta(t') dt'\right]\right\}. \tag{33}$$

For the pulsed gradient spin echo (PGSE) and the pulsed gradient stimulated-echo (PGSTE) experiments [17,20,21] as shown in Fig. 1, Eq. (33) can be easily integrated as

$$A(t) = \exp\left[-\sum_{i=x',y',z'} D'_{f_1 \text{or} 2 ii} (\gamma g_i' \delta)^{\beta_i} (\Delta - \frac{\beta_i - 1}{\beta_i + 1} \delta)\right], \tag{34}$$

which reproduces the result obtained by modified Bloch equations [40,41].

2.3. Restricted time fractional diffusion

Restricted isotropic time-fractional diffusion has been studied in several reports [47, 48, 49]. The PFG signal attenuation expressions of restricted isotropic fractional diffusion have been obtained in different geometries such as plate, sphere and cylinder [35]. These restricted fractional diffusions could exist in confined polymer systems. Many polymer systems have anisotropic behavior. Hence, it may be necessary to develop some theory treatments for PFG signal attenuation of restricted anisotropic anomalous diffusion; furthermore, a good starting point may be to investigate the simple restricted anisotropic fractional diffusion, diffusion within a box with reflecting boundaries.

Based on fractal derivative, the anisotropic time-fractional diffusion may be described by

$$\frac{\partial}{\partial t^\alpha} P' = D'_{f_1 x' x'} \frac{\partial^2}{\partial x'^2} P' + D'_{f_1 y' y'} \frac{\partial^2}{\partial y'^2} P' + D'_{f_1 z' z'} \frac{\partial^2}{\partial z'^2} P', \quad (35)$$

while from fractional derivative, it may be described by

$$_t D_*^\alpha P' = D'_{f_1 x' x'} \frac{\partial^2}{\partial x'^2} P' + D'_{f_1 y' y'} \frac{\partial^2}{\partial y'^2} P' + D'_{f_1 z' z'} \frac{\partial^2}{\partial z'^2} P'. \quad (36)$$

The reflecting boundary condition for Eqs. (35) and (36) is

$$\begin{cases} P(x', y', z', 0) = \delta(x')\delta(y')\delta(z') \\ \frac{\partial P(x', t)}{\partial x} =, x' = \pm \frac{a_{x'}}{2} \\ \frac{\partial P(y', t)}{\partial y'} = 0, y' = \pm \frac{a_{y'}}{2}, \\ \frac{\partial P(z', t)}{\partial z'} = 0, z' = \pm \frac{a_{z'}}{2} \end{cases} \quad (37)$$

where $a_i, i = x', y', z'$ is the distance between the parallel surfaces of the box along $i$ direction in the principal frame. The separation of variables method can be used to solve the time-fractional diffusion equations [1,2,35,46]. The probability density function by separating variable method can be set as

$$P' = \sum_{l=0}^{\infty} \sum_{m=0}^{\infty} \sum_{n=0}^{\infty} r_l(x') r_m(y') r_n(z') T_{l,m,n}(t), \quad (38)$$

where $r_l(x') r_m(y') r_n(z')$ is the spatial function and has been obtained in Ref. [46], which is

$$r_l(x') r_m(y') r_n(z') = \frac{2^{(\varepsilon_{l0} + \varepsilon_{m0} + \varepsilon_{n0})}}{a_{x'} a_{y'} a_{z'}} \cos(\frac{l\pi}{a_{x'}} x') \cos(\frac{l\pi}{a_{x'}} x'_0) \cos(\frac{m\pi}{a_{y'}} y') \cos(\frac{m\pi}{a_{y'}} y'_0) \cos(\frac{n\pi}{a_{z'}} z') \cos(\frac{n\pi}{a_{z'}} z'_0)$$

, (39)

where $\varepsilon_{i0} = \begin{cases} 1, i \neq 0 \\ 0, i = 0 \end{cases}$, and

$$T_{l,m,n}(t) = \begin{cases} \exp\left\{-\left[D_{f_1 x' x'}\left(\frac{l\pi}{a_{x'}}\right)^2 + D_{f_1 y' y'}\left(\frac{m\pi}{a_{y'}}\right)^2 + D_{f_1 z' z'}\left(\frac{n\pi}{a_{z'}}\right)^2\right] t^\alpha\right\}, \text{fractal derivative} \\ E_{\alpha,1}\left\{-\left[D_{f_2 x' x'}\left(\frac{l\pi}{a_{x'}}\right)^2 + D_{f_2 y' y'}\left(\frac{m\pi}{a_{y'}}\right)^2 + D_{f_2 z' z'}\left(\frac{n\pi}{a_{z'}}\right)^2\right] t^\alpha\right\}, \text{fractional derivative} \end{cases} \quad (40)$$

The PFG signal attenuation for PGSE or PGSTE experiments will be

$$A(t) = \int_{-\infty}^{\infty} P' \exp(-i\gamma \delta \mathbf{g}' \cdot \mathbf{r}') dV' = (\gamma g'_{x'} \delta a_{x'} \gamma g'_{y'} \delta a_{y'} \gamma g'_{z'} \delta a_{z'})^2 \sum_{l=0}^{\infty} \sum_{m=0}^{\infty} \sum_{n=0}^{\infty} T_{l,m,n}(t)(h_l h_m h_n)$$

$$\times \frac{1-(-1)^l \cos(\gamma g'_{x'} \delta a_{x'})}{[(\gamma g'_{x'} \delta a_{x'})^2 - (l\pi)^2]^2} \times \frac{1-(-1)^m \cos(\gamma g'_{y'} \delta a_{y'})}{[(\gamma g'_{y'} \delta a_{y'})^2 - (m\pi)^2]^2} \times \frac{1-(-1)^n \cos(\gamma g'_{z'} \delta a_{z'})}{[(\gamma g'_{z'} \delta a_{z'})^2 - (n\pi)^2]^2} ,$$

(41)

where $h_i = \begin{cases} 2, i = 0 \\ 4, i \neq 0 \end{cases}$. Eq. (41) can be reduced to the reported one-dimensional restricted time-fractional diffusion, namely restricted anomalous diffusion within a plate in Ref. [35]. In restricted time-fractional diffusion, the spatial dependence of the net signal attenuation is the same as that in normal diffusion, which is very reasonable as $\beta = 2$ for both normal and time-fractional diffusion. When $\alpha = 1, \beta = 2$, the Mittag-Leffler function reduces to an exponential function, and restricted fractional reduces to restricted normal diffusion.

2.4 Anisotropic curvilinear diffusion

In the above, both the PFG signal attenuation expressions for free and restricted anisotropic fractional diffusions were derived, which may be abstract. In this part, more realistic examples, the PFG curvilinear diffusions will be analyzed to help our understanding of PFG anisotropic fractional diffusion. In polymer systems, the curvilinear diffusion could be either the Brownian motion of chain segments or a small penetrant (molecule or ion) diffusing along a polymer chain [50]. The free and restricted isotropic curvilinear diffusions have been analyzed by traditional methods in Refs. [4,23,35]. The similar method will be used to analyze PFG anisotropic curvilinear diffusion, which may result from anisotropic chain structure or anisotropic penetrant diffusion constant. For simplicity, penetrant with an isotropic diffusion constant diffusing along an anisotropic chain with three different segment lengths $\zeta_i, i = x', y', z'$ will be considered here.

In polymer or biological system, a small molecule or ion may prefer to move along a certain curvilinear path. For example, in a blend of poly(methyl methacrylate) (PMMA) and poly(ethylene oxide) (PEO), PEO chain provides a fast path for diethyl ether to diffuse [51]; in PEO-based polymer electrolytes for rechargeable batteries, the lithium ion hops and 'slides' along PEO backbone [52]. When a small penetrant molecule diffuses along an ideal and infinitely long polymer chain (for simplicity, it is assumed to be immobile as the polymer chain motion is much slower than penetrant), at time $\Delta$, the penetrant distribution probability on the chain can be described as

$$\frac{1}{\sqrt{4\pi D_0 \Delta}} \exp\left(-\frac{(N\bar{\zeta})^2}{4D_0 \Delta}\right),$$ 
(42)

where $N$ is the number of segments away from the starting segment, $\bar{\zeta} = \sum_{i=x',y',z'} \zeta_i$ is the average segment length (it is safe to assume that $N_i = N/3, i = x', y', z'$ when $N$ is big enough.), and $D_0$ is the local penetrant diffusion constant along the chain without being affected by the chain folding [35]. The differences of segment length among different directions could result from stretching or compressing of polymer material, which could lead to an anisotropic diffusion (Other possible mechanisms may be proposed for anisotropic diffusion). The probability distribution for the end to end distance of a freely folding chain can be viewed as a result of a virtual diffusion [35], which has jump lengths $\zeta_i$, $i = x', y', z'$ and a jump time $t_{jump}$ arbitrarily set to one. The virtual diffusion coefficient $D_{sii}$ and the diffusion time $\tau$ will be

$$\begin{cases} D_{sii} = \dfrac{\zeta_i^2}{6t_{jump}} = \dfrac{\zeta_i^2}{6} \\ \tau = Nt_{jump} = N \approx 3Nz \\ \overline{\zeta} = \sum_{i=x',y',z'} \zeta_i \end{cases} \qquad (43)$$

Such a virtual free normal diffusion has an anisotropic Gaussian probability density function [44]

$$P' = \frac{1}{(4\pi N)^{3/2} \sqrt{D'_{sx'x'} D'_{sy'y'} D'_{sz'z'}}} \exp\left[-\frac{\xi_i^2}{4D'_{sx'x'}N} - \frac{\xi_j^2}{4D'_{sx'x'}N} - \frac{\xi_k^2}{4D'_{sx'x'}N}\right]$$
$$= \frac{1}{(2\pi N/3)^{3/2} \sqrt{\zeta_i \zeta_j \zeta_k}} \exp\left[-\frac{\xi_i^2}{2N\zeta_i/3} - \frac{\xi_j^2}{2N\zeta_j/3} - \frac{\xi_k^2}{2N\zeta_k/3}\right] \qquad (44)$$

The PFG signal attenuation for free curvilinear diffusion will be

$$A(\Delta) = 2\int_0^\infty \frac{1}{\sqrt{4\pi D_0 \Delta}} \exp\left(-\frac{(N\overline{\zeta})^2}{4D_0\Delta}\right)$$
$$\times \left[2\int_0^\infty \frac{1}{(2\pi N/3)^{3/2}\sqrt{\zeta_i \zeta_j \zeta_k}} \exp\left[-\frac{\xi_i^2}{2N\zeta_i/3} - \frac{\xi_j^2}{2N\zeta_j/3} - \frac{\xi_k^2}{2N\zeta_k/3}\right] \cos(\sum_i rg'_i \delta \xi_i) d\xi_i\right] dN\overline{\zeta}$$
$$= 2\int_0^\infty \frac{1}{\sqrt{4\pi D_0 \Delta}} \exp\left(-\frac{(N\overline{\zeta})^2}{4D_0\Delta}\right) \exp\left[-\sum_i (rg'_i\delta)^2 N\zeta_i^2/6\right] dN\overline{\zeta} \qquad (45)$$
$$= \exp(\sum_i \frac{\sqrt{D_0\Delta}\gamma^2 g_i'^2 \delta^2 \zeta_i^2}{3\overline{\zeta}})^2 \, erfc(\sum_i \frac{\sqrt{D_0\Delta}\gamma^2 g_i'^2 \delta^2 \zeta_i^2}{3\overline{\zeta}})$$
$$= E_{1/2,1}(-\sum_i \frac{\sqrt{D_0\Delta}\gamma^2 g_i'^2 \delta^2 \zeta_i^2}{3\overline{\zeta}})$$

From Eq. (45), we get

$$D_{f_2 ii} = \frac{\sqrt{D_0\Delta}\zeta_i^2}{3\overline{\zeta}}, i = x', y', z'. \qquad (46)$$

If the anisotropic freely folding chain is confined in a specific structure, the probability distribution function can be obtained similarly by a restricted virtual diffusion with some approximations: the excluded volume effect [53] is neglected and the two segments meet and reflect at the same point of the boundary are still connected rather than broken. The anisotropic restricted virtual diffusion has $D_{sii} = \dfrac{\zeta_i^2}{6}$ and diffusion time $\tau = N$, which are the same as that in free virtual diffusion. The possibility distribution function $P_s(\mathbf{r}|\mathbf{r}',\tau)$ can be obtained by eigenfunction expansion [1,2,35]

$$P_s(\mathbf{r}|\mathbf{r}',\tau) = \sum_{l=0}^\infty \sum_{m=0}^\infty \sum_{n=0}^\infty r_l(x') r_m(y') r_n(z') \exp(-\lambda_l D_{sx'x'}\tau - \lambda_m D_{sy'y'}\tau - \lambda_n D_{sz'z'}\tau)$$
$$= \sum_{l=0}^\infty \sum_{m=0}^\infty \sum_{n=0}^\infty r_l(x') r_m(y') r_n(z') \exp(-\lambda_l \frac{N\zeta_{x'}^2}{6} - \lambda_m \frac{N\zeta_{y'}^2}{6} - \lambda_n \frac{N\zeta_{z'}^2}{6}),$$
$$\qquad (47)$$

where $r_l(x')r_m(y')r_n(z')$ is the spatial function for restricted anisotropic normal diffusion. The penetrant probability distribution function will be

$$P(\mathbf{r}|\mathbf{r}',\Delta) = 2\int_0^\infty \frac{1}{\sqrt{4\pi D_0 \Delta}} \exp\left(-\frac{(N\overline{\zeta})^2}{4D_0\Delta}\right)[P_s(\mathbf{r}|\mathbf{r}',\tau)]dN\overline{\zeta}$$

$$= \sum_{l,m,n} r_l(x')r_m(y')r_n(z') 2\int_0^\infty \frac{1}{\sqrt{4\pi D_0 \Delta}} \exp\left(-\frac{(N_z\overline{\zeta})^2}{4D_0\Delta}\right)\exp(-\lambda_l \frac{N\zeta_{x'}^2}{6} - \lambda_m \frac{N\zeta_{y'}^2}{6} - \lambda_n \frac{N\zeta_{z'}^2}{6})dN\overline{\zeta}$$

$$= \sum_{l,m,n} r_l(x')r_m(y')r_n(z')\exp(\frac{\sqrt{D_0\Delta}(\lambda_l\zeta_{x'}^2 - \lambda_m\zeta_{y'}^2 - \lambda_n\zeta_{z'}^2)}{3\overline{\zeta}})^2 erfc(\frac{\sqrt{D_0\Delta}(\lambda_l\zeta_{x'}^2 - \lambda_m\zeta_{y'}^2 - \lambda_n\zeta_{z'}^2)}{3\overline{\zeta}})$$

$$= \sum_{l,m,n} r_l(x')r_m(y')r_n(z') E_{1/2,1}(-\frac{\sqrt{D_0\Delta}(\lambda_l\zeta_{x'}^2 - \lambda_m\zeta_{y'}^2 - \lambda_n\zeta_{z'}^2)}{3\overline{\zeta}})$$

(48)

Again, we have

$$D_{f_2 ii} = \frac{\sqrt{D_0\Delta}\zeta_i^2}{3\overline{\zeta}}, i = x', y', z',$$ (49)

which is the same as that in Eq. (46). The PFG signal attenuation of a penetrant diffusion in a confined polymer structure will be

$$A_{SGP}(\Delta) = \iint_{\mathbf{r},\mathbf{r}'} \rho(\mathbf{r},0) P(\mathbf{r}|\mathbf{r}',\Delta) \exp(i\mathbf{K}(\delta)\cdot(\mathbf{r}'-\mathbf{r}))d\mathbf{r}d\mathbf{r}'$$

$$= \sum_{l=0}^\infty \sum_{m=0}^\infty \sum_{n=0}^\infty E_{1/2,1}(-\frac{\sqrt{D_0\Delta}(\lambda_l\zeta_{x'}^2 - \lambda_m\zeta_{y'}^2 - \lambda_n\zeta_{z'}^2)}{3\overline{\zeta}})$$

$$\times \iint_{\mathbf{r},\mathbf{r}'} \rho(\mathbf{r},0) r_l(x')r_m(y')r_n(z') \exp(i\mathbf{K}(\delta)\cdot(\mathbf{r}'-\mathbf{r}))d\mathbf{r}d\mathbf{r}'$$

$$= \sum_{n=0} T_{l,m,n,f_2}(\Delta) X_l(x') X_m(y') X_n(z')$$

(50)

where

$$T_{l,m,n,f_2}(\Delta) = E_{1/2,1}(-\frac{\sqrt{D_0\Delta}(\lambda_l\zeta_{x'}^2 - \lambda_m\zeta_{y'}^2 - \lambda_n\zeta_{z'}^2)}{3\overline{\zeta}}),$$ (51)

and

$$X_l(x')X_m(y')X_n(z') = \iint_{\mathbf{r},\mathbf{r}'} \rho(\mathbf{r},0) r_l(x')r_m(y')r_n(z') \exp(i\mathbf{K}(\delta)\cdot(\mathbf{r}'-\mathbf{r}))d\mathbf{r}d\mathbf{r}'.$$ (52)

Eq. (50) is a specific case of restricted anisotropic fractional diffusion based on the fractional derivative model with $\alpha = 1/2$ and $D_{f_2 ii} = \frac{\sqrt{D_0\Delta}\zeta_i^2}{3\overline{\zeta}}, i = x', y', z'$. Thus, the PFG curvilinear diffusion results obtained from the conventional methods of normal diffusion agree with the results from the PFG fractional diffusion based on the fractional derivative model. The above derivation is based on an anisotropic Gaussian diffusion along an ideal anisotropic Gaussian chain. In real polymer systems, many factors such as the excluded volume effect, the structure and size of the chain, and the hopping of penetrant between different chains will affect the diffusion and could make the time and space derivative parameters deviate from $\alpha = 1/2$ and $\beta = 2$ for a Gaussian diffusion along an ideal Gaussian chain.

## 4. Results and Discussion

**Table 1.** Comparison of the signal attenuation of anisotropic anomalous diffusion obtained from the ISA method with that from other methods.

| General Fractional diffusion $\{0 < \alpha, \beta \leq 2\}$ |
|---|
| Fractal derivative: $\exp(-\mathbf{b}'^*:\mathbf{D}'_{f_1})$.     Fractional derivative: $E_{\alpha,1}(-\mathbf{b}'^*:\mathbf{D}'_{f_2})$. <br><br> $\mathbf{b}'^* = \int_0^t \widetilde{\mathbf{K}}^*(\mathbf{t}') \otimes \mathbf{K}^*(\mathbf{t}') dt'^\alpha$, $\mathbf{b}^* = \int_0^t [\widetilde{\mathbf{K}}^*(\mathbf{t}') \cdot R] \otimes [\widetilde{R} \cdot \mathbf{K}^*(\mathbf{t}')] dt'^\alpha$, $\mathbf{K}^*(\mathbf{t}') = \begin{pmatrix} K'^{\beta_{x'}/2}_{x'}(t)\mathbf{i}_{x'} \\ K'^{\beta_{y'}/2}_{y'}(t)\mathbf{i}_{y'} \\ K'^{\beta_{z'}/2}_{z'}(t)\mathbf{i}_{z'} \end{pmatrix}$, $\mathbf{K}'(\mathbf{t}') = R\mathbf{K}(\mathbf{t}')$, <br><br> $-\mathbf{b}'^*:\mathbf{D}'_{f_j} = -\mathbf{b}^*:\mathbf{D}_{f_j} = \sum_{i=x',y',z'} \int_0^t D'_{f_jii} K'^{\beta_i}_i(t') dt'^\alpha$ <br><br> $= \sum_{i=x',y',z'} D'_{f_jii} \gamma^{\beta_i} g'^{\beta_i}_i \left\{ \frac{\alpha}{\beta_i + \alpha} \delta^{\beta_i + \alpha} + \delta^{\beta_i}(\Delta^\alpha - \delta^\alpha) + \alpha(\Delta + \delta)^{\beta_i(\alpha+1)} \left[ B(\alpha, \beta_i + 1) - B(\frac{\Delta}{\Delta + \delta}; \alpha, \beta_i + 1) \right] \right\}$ **a,b <br><br> $\xrightarrow{SGP} \approx \sum_{i=x',y',z'} D'_{f_jii} \gamma^{\beta_i} g'^{\beta_i}_i \Delta^\alpha$ |
| **a,b the integration has been obtained in Ref. [16] for isotropic anomalous diffusion; $f_j$ stands for $f_1$ or $f_2$; B(x,y) and B(x;a,b) are Beta function and incomplete Beta function, respectively. |
| Other methods: <br> $E_{\beta, 1+\alpha/\beta, \alpha/\beta}(-\gamma^\alpha t^{\alpha+\beta} \int_g |\mathbf{g}.\mathbf{y}|^\alpha m(d\mathbf{y}))$, from Hanyga et al. [29] |
| Space-fractional diffusion $\{\alpha = 1, 0 < \beta \leq 2\}$ |
| Fractal derivative: $A(t) = \exp\left[-\sum_{i=x',y',z'} D'_{f_1ii} (\gamma g'_i \delta)^{\beta_i} (\Delta - \frac{\beta_i - 1}{\beta_i + 1}\delta)\right]$. <br><br> Fractional derivative: $A(t) = \exp\left[-\sum_{i=x',y',z'} D'_{f_2ii} (\gamma g'_i \delta)^{\beta_i} (\Delta - \frac{\beta_i - 1}{\beta_i + 1}\delta)\right]$. |
| Other methods: <br> $\exp\left[-\sum_{i=x',y',z'} D'_{fii} \mu^{2(\beta^*_i/2-1)} (\gamma g'_i \delta)^{\beta_i} (\Delta - \frac{\beta_i - 1}{\beta_i + 1}\delta)\right]$ [40] |
| Time-fractional diffusion $\{0 < \alpha \leq 2, \beta = 2\}$ |
| $A(t) = \begin{cases} \exp\left\{-\left[\sum_{i=x',y',z'} D'_{f_1ii} \int_0^t k'^2_i(t') dt'^\alpha\right]\right\} = \exp(-\mathbf{b}'^*_t:\mathbf{D}'_{f_1}) = \exp(-\mathbf{b}_t^*:\mathbf{D}_{f_1}), & \textit{fractal derivative} \\ E_{\alpha,1}\left\{-\left[\sum_{i=x',y',z'} D'_{f_2ii} \int_0^t k'^2_i(t') dt'^\alpha\right]\right\} = E_{\alpha,1}(-\mathbf{b}'^*_t:\mathbf{D}'_{f_2}) = E_{\alpha,1}(-\mathbf{b}_t^*:\mathbf{D}_{f_2}), & \textit{fractional derivative} \end{cases}$ <br><br> $\sum_{i=x',y',z'} D'_{f_jii} \int_0^t k'^2_i(t') dt'^\alpha$ <br><br> $= \sum_{i=x',y',z'} D_{f_jii} \gamma^2 g'^2_i \left\{ \frac{\alpha \delta^{\alpha+2}}{\alpha + 2} + \delta^2(\Delta^\alpha - \delta^\alpha) + \frac{2}{(\alpha+1)(\alpha+2)}\left[(\Delta+\delta)^{2+\alpha} - \Delta^{2+\alpha}\right] - \frac{2}{(\alpha+1)}\Delta^{1+\alpha}\delta - \Delta^\alpha \delta^2 \right\}$ <br><br> $= \begin{cases} \sum_{i=x',y',z'} D_{f_jii} \gamma^2 g'^2_i \delta^2 (\Delta - \delta/3), \alpha = 1 \\ \sum_{i=x',y',z'} D_{f_jii} \gamma^2 g'^2_i \delta^{\alpha+2} \frac{4(2^\alpha - \alpha - 2)}{(\alpha+1)(\alpha+2)}, \Delta = \delta \\ \sum_{i=x',y',z'} D_{f_jii} \gamma^2 g'^2_i \delta^2 \Delta^\alpha, \textit{if } \delta \textit{ is neglected.} \end{cases}$ |

Other methods
1. $\exp\left[-D\gamma^2 g^2 \delta^2 (\Delta - \delta/3)\right]$, free normal diffusion [20,21]

2. $A(t) = \exp\left[-D_{f_1}\int_0^t K^\beta(t')dt'^\alpha\right]$, $A(t) = E_{\alpha,1}\left[-D_{f_2}\int_0^t K^\beta(t')dt'^\alpha\right]$ [16,32]

3. $A(t) = \exp(-\mathbf{b}':\mathbf{D}') = \exp(-\mathbf{b}:\mathbf{D})$ [42]

4. Free and restricted curvilinear diffusion derived in this paper:
$$A_{SGP,free}(\Delta) = \exp(\sum_i \frac{\sqrt{D_0\Delta}\gamma^2 g_i'^2 \delta^2 \zeta_i^2}{3\overline{\zeta}})^2 \, erfc(\sum_i \frac{\sqrt{D_0\Delta}\gamma^2 g_i'^2 \delta^2 \zeta_i^2}{3\overline{\zeta}}) = E_{1/2,1}(-\sum_i \frac{\sqrt{D_0\Delta}\gamma^2 g_i'^2 \delta^2 \zeta_i^2}{3\overline{\zeta}});$$

$$A_{SGP,restricted}(\Delta) = \sum_{l=0}^\infty \sum_{m=0}^\infty \sum_{n=0}^\infty E_{1/2,1}\left(-\frac{\sqrt{D_0\Delta}(\lambda_l \zeta_{x'}^2 - \lambda_m \zeta_{y'}^2 - \lambda_n \zeta_{z'}^2)}{3\overline{\zeta}}\right) X_l(x') X_m(y') X_n(z')$$

$$D_{f_2 ii} = \frac{\sqrt{D_0\Delta}\zeta_i^2}{3\overline{\zeta}}, \, i = x', y', z'$$

The general PFG signal attenuation expressions for free anisotropic anomalous diffusion were obtained by the instantaneous signal attenuation method based on the fractal and fractional derivative models. The method was also used to investigate the restricted anisotropic anomalous diffusion in a box structure. Some of the obtained results were summarized and compared with these obtained from other methods in Table 1. From Table 1, the anisotropic anomalous diffusion result can be reduced to reported anisotropic normal diffusion result when time derivative order $\alpha = 1$ and space derivative order $\beta = 2$. The anisotropic anomalous diffusion result can also be reduced to reported isotropic anomalous diffusion result when $D'_{f_1 x'x'} = D'_{f_1 y'y'} = D'_{f_1 z'z'} = D_{f_1}$ and $\beta_{x'} = \beta_{y'} = \beta_{z'} = \beta$. The results of space-fractional diffusion agree with that obtained by the modified Bloch equation method [40]. Fig. 2 demonstrates different types of anisotropic anomalous diffusions and their signal attenuation expressions. The anisotropic normal diffusion is a single point in this figure and may be view as a specific case of the anisotropic fractional diffusion.

The two different types of attenuations, stretched exponential function (SEF) and Mittag-Leffler function (MLF) attenuations, have similarities and differences as shown in Fig. 3. At small signal attenuation, the MLF can be approximated as an SEF, while at large signal attenuation, they are significantly different. For subdiffusion as shown in Fig. 3a, $E_{\alpha,1}(-x) > \exp(-x), x > 0$, which means that the MLF attenuates more slowly than the SEF. While, for superdiffusion as shown in Fig 3b, $E_{\alpha,1}(-x) < \exp(-x)$ at small $x$ values, then $E_{\alpha,1}(-x)$ oscillates between positive and negative values at large $x$ values, which still has no clear explanation now. It may be due to the fat-tailed probability density function. Fig. 3c and 3d show that the dependences of signal attenuation plotted in logarithm scale upon $g^2$ are different between the Mittag-Leffler function attenuation and the stretched exponential attenuation. The Mittag-Leffler function attenuation is nonlinearly dependent upon $g^\beta$ while the stretched exponential is linearly dependent upon $g^\beta$.

In section 2.4, the derived PFG signal attenuation expressions of anisotropic curvilinear diffusion based on conventional method [4] agree with the PFG signal attenuation of anisotropic anomalous diffusion based on the fractional derivative model. The PFG anisotropic curvilinear diffusion can be viewed as a specific case of anisotropic anomalous diffusion with $\alpha = 1/2$ and $\beta = 2$, which is like that in isotropic curvilinear diffusion reported in Ref. [35]. For a non-ideal curvilinear diffusion, the penetrant diffusion may not be Gaussian, and the polymer chain may not be a Gaussian chain. Additionally other complications such as excluded volume effect, penetrant hopping between different polymer chain may make the curvilinear diffusion have $\alpha \neq 1/2$ and $\beta \neq 2$.

Under SGP approximation, the PFG signal attenuation for PGSE and PGSTE experiments from Eqs (11) and (25) is

$$A_{SGP}(t) = \begin{cases} \exp\left[-\sum_{i=x',y',z'} D'_{f_1 ii}(\gamma g'_i \delta)^{\beta_i} t^\alpha\right], & \text{fractal derivative} \\ E_{\alpha,1}\left[-\sum_{i=x',y',z'} D'_{f_1 ii}(\gamma g'_i \delta)^{\beta_i} t^\alpha\right], & \text{fractional derivative} \end{cases} \quad (53)$$

In the real material, the anisotropic anomalous diffusion may have $D'_{fx'x'} = D'_{fy'y'} = D_{f\perp}$ and $D'_{fz'z'} = D'_{f\parallel}$. If the gradient field is applied in only one direction such as $z$ direction and the angle between the $z$ direction and the principal axes $z'$ is $\theta$, from Eq. (53), the PFG signal attenuation will be

$$A_{SGP}(t,\theta) = \begin{cases} \exp\left\{-(\gamma g \delta)^{\beta_\perp} t^\alpha \left[D'_{f\parallel} \frac{(\gamma g \delta)^{\beta_\parallel}}{(\gamma g \delta)^{\beta_\perp}} \cos^2\theta + D'_{f\perp} \sin^2\theta\right]\right\}, & \text{fractal derivative} \\ E_{\alpha,1}\left\{-(\gamma g \delta)^{\beta_\perp} t^\alpha \left[D'_{f\parallel} \frac{(\gamma g \delta)^{\beta_\parallel}}{(\gamma g \delta)^{\beta_\perp}} \cos^2\theta + D'_{f\perp} \sin^2\theta\right]\right\}, & \text{fractional derivative} \end{cases} \quad (54)$$

The averaging over all angles will give [20]

$$A_{SGP}(t) = \frac{\int_0^\pi A_{SGP}(t,\theta)\sin\theta d\theta}{\int_0^\pi \sin\theta d\theta} . \quad (55)$$

The similar expression to Eq. (55) has been reported in anisotropic normal diffusion in Ref. [20]. From the calculation result of Ref. [20], for the fractal derivative model, Eq. (55) can be further simplified as

$$A_{SGP}(t,\theta) = \exp\left[-(\gamma g \delta)^{\beta_\perp} D'_{f\perp} t^\alpha\right] \int_0^1 \exp\left\{-(\gamma g \delta)^{\beta_\perp} t^\alpha \left[D'_{f\parallel} \frac{(\gamma g \delta)^{\beta_\parallel}}{(\gamma g \delta)^{\beta_\perp}} - D'_{f\perp}\right]\right\} x^2 dx . \quad (56)$$

Similarly to anisotropic normal diffusion, the Eqs. (54)-(56) can be used to determine the dimension of the structure. For instance, if penetrant molecules such as water or small organic molecules diffusing in some materials are investigated, the symmetry of materials could be a "one-dimensional" or "capillary" and $D_{f_{1\perp}} \ll D'_{f_{1\parallel}}$, the attenuation will be

$$A_{SGP,1D}(t,\theta) = \int_0^1 \exp\left[-(\gamma g \delta)^{\beta_\parallel} D'_{f_1\parallel} t^\alpha\right] x^2 dx. \tag{57}$$

While, if the symmetry is a "two-dimensional" or "lamellar" and $D_{f_1\perp} \gg D'_{f_1\parallel}$, the attenuation will be

$$A_{SGP,2D}(t) = \exp\left[-(\gamma g \delta)^{\beta_\perp} D'_{f_1\perp} t^\alpha\right]\int_0^1 \exp\left[-(\gamma g \delta)^{\beta_\perp} t^\alpha D_{f_1\perp}\right] x^2 dx. \tag{58}$$

The curves of ($A_{SGP}(t)$) plotted on a logarithmic scale versus $(\gamma g \delta)^{\beta_\perp} t^\alpha$ will have significant difference among $A_{SGP,1D}(t)$, $A_{SGP,2D}(t)$, and $A_{SGP,3D}(t)$ as shown in Fig. 4. From Fig. 4, the curve of the one-dimensional anisotropic fractional diffusion signal attenuation $A_{SGP,1D}(t)$ plotted on a logarithmic scale has biggest bending among the three curves $A_{SGP,1D}(t)$, $A_{SGP,2D}(t)$, and $A_{SGP,3D}(t)$. The curve of $A_{SGP,3D}(t)$ from the fractal derivative model plotted on a logarithmic scale has a linearly dependent upon $(\gamma g \delta)^{\beta_\perp} t^\alpha$. Again, the Mittag-Leffler function attenuation is different from the stretched exponential attenuation in Fig. 4.

For the restricted anomalous diffusion, the ISA method is very simple. The other methods for PFG normal restricted diffusion may be extended to study restricted anomalous diffusion. For instance, the multiple-narrow-pulse approximation [54] method, which is a different kind of propagator method that can be described as

$$A(t) = \iiint_j \exp\left[i \cdot \left(\sum_j^{2n+1} 2\pi\gamma g(t_j) dt_j x(t_j)\right)\right] \cdot \prod_{j=1}^{2n} P_j\left(x(t_{j+1}),(t_{j+1}-t_j)|x(t_j)\right) dx(t_1)\cdots ddx(t_{2n+1}), \tag{59}$$

where $P_j\left(x(t_{j+1}),(t_{j+1}-t_j)|x(t_j)\right)$ is the propagator for a particle moving from $x(t_j)$ to $x(t_{j+1})$ during $t_j$ to $t_{j+1}$, and the total time is divided by 2n+1 periods including 2n small gradient impulses and the diffusion delay. The multiple-narrow-pulse approximation method [54] calculates the total signal attenuation by averaging over all possible phase pathways describing by $\sum_j^{2n+1} 2\pi\gamma g(t_j) dt_j x(t_j)$, which is significantly different propagator method from the ISA method in which the total signal attenuation is the accumulation of the instantaneous signal attenuation in each small interval. The evaluation of Eq. (59) is more complicated, but should provide an important way to calculate the PFG restricted anomalous diffusion, which will be studied in future.

The ISA method is simple. The fundamental basis of ISA method is that diffusion is a cumulative process. For free normal diffusion, the cumulative process can be viewed as a cumulative product of instantaneous signal attenuation of each interval. For free anomalous diffusion, the change of the PFG signal intensity after a specific interval could be increasing rather than attenuating such as that in a certain situation of supper diffusion shown in Fig. 3b, therefore, the "attenuation" in the name of instantaneous signal attenuation method may lead to confusion. The ISA method may be called as diffusion cumulative process method. Under SGP approximation, such a cumulative process gives the same signal attenuation expressions as that obtained by SGP approximation. When the FGPW effect is considered, for free normal diffusion, the ISA method reproduces the well-known result $\exp\left[-\sum_i D_i \gamma^2 g_i'^2 \delta^2 (\Delta - \delta/3)\right]$. While, for free and restricted anomalous diffusion, at small $\delta/\Delta$, the ISA method should be a good approximation based on the comparison with 1D CTRW simulation [32]. The 3D CTRW simulation will be performed in future.

The PFG anisotropic anomalous diffusion results provide new formalisms for PFG anomalous diffusion studies in NMR and MRI, which can potentially be applied in Diffusion tensor imaging (DTI) [55]. The DTI, has been extensively used in MRI for studying brain tissues, muscles and so on, and the current DTI studies are based on PFG anisotropic or isotropic normal diffusion [54]. The introducing of free and restricted anisotropic anomalous diffusions could provide additional insights in those studies.

**Appendix A. Definition of fractal derivative from reference [11-12].**

$$\frac{\partial P^{\beta/2}}{\partial t^\alpha} = \lim_{t_1 \to t} \frac{P^{\beta/2}(t_1) - P^{\beta/2}(t)}{t_1^\alpha - t^\alpha}, 0 < \alpha, 0 < \beta/2 . \tag{A.1}$$

**Appendix B. Definition of fractional derivative [3,13-15].**

$$\frac{d^\beta}{d|z|^\beta} = -\frac{1}{2\cos\frac{\pi\alpha}{2}}\left[{}_{-\infty}\boldsymbol{D}_z^\beta + {}_z\boldsymbol{D}_\infty^\beta\right],$$

where

$$_{-\infty}\boldsymbol{D}_z^\beta f(z) = \frac{1}{\Gamma(m-\beta)} \frac{d^m}{dz^m} \int_{-\infty}^z \frac{f(y)dy}{(z-y)^{\beta+1-m}}, \beta > 0, m-1 < \beta < m , \tag{B.1}$$

and

$$_z\boldsymbol{D}_\infty^\beta f(z) = \frac{(-1)^m}{\Gamma(m-\beta)} \frac{d^m}{dz^m} \int_z^\infty \frac{f(y)dy}{(y-z)^{\beta+1-m}}, \beta > 0, m-1 < \beta < m . \tag{B.2}$$

**Figure Legends**

**Fig. 1** (a) PGSE pulse sequences with gradient pulses of finite length $\delta$, (b) PGSTE pulse sequence with gradient pulse of finite length $\delta$.

**Fig. 2.** The dependence of signal attenuations of different types of PFG anisotropic fractional diffusions upon the time and space derivative parameters $\alpha$ and $\beta$.

**Fig. 3.** Comparison Mittag-Leffler function attenuation with stretched exponential attenuation: (a) signal attenuation versus $\Delta^\alpha$ for subdiffusion with $\alpha = 0.8, \beta = 2$, $g'_{x'} = g'_{y'} = 0$ and $g'_{z'} =30$ gauss/cm, (b) signal attenuation versus $\Delta^\alpha$ for superdiffusion with $\alpha =1.5, \beta = 2$, $g'_{x'} = g'_{y'} = 0$ and $g'_{z'} =30$ gauss/cm, (c) signal attenuation versus $g^2$ at small $\delta/\Delta$ for subdiffusion with $\alpha = 0.8, \beta = 2$, $g'_{x'} = g'_{y'} = 0$ and varying $g'_{z'}$, (d) signal attenuation versus $g^2$ at big $\delta/\Delta$ for subdiffusion with $\alpha = 0.8, \beta = 2$, $g'_{x'} = g'_{y'} = 0$ and varying $g'_{z'}$. The diffusion coefficients are $D'_{f_1 x'x'} = D'_{f_1 y'y'} = D'_{f_1 z'z'}$ 1.38e-9 × $10^{-9}$ m$^\beta$/s$^\alpha$.

**Fig. 4.** Comparison PFG signal attenuation of 1D, 2D and 3D anisotropic fractional diffusion: (a) stretched exponential attenuation, the values for ($D_{f_1\parallel}$, $D_{f_1\perp}$) are ($3.5 D_{f_1 0}$, $10^6 D_{f_1 0}$), ($10^6 D_{f_1 0}$, $D_{f_1 0}$) and ($0.5 D_{f_1 0}$, $0.5 D_{f_1 0}$) for 1D, 2D and 3D respectively, (b) Mittag-Leffler function attenuation, the values for ($D_{f_1\parallel}$, $D_{f_1\perp}$) are ($3.5 D_{f_2 0}$, $10^{-6} D_{f_2 0}$), ($10^6 D_{f_2 0}$, $D_{f_2 0}$) and ($0.5 D_{f_2 0}$, $0.5 D_{f_2 0}$) for 1D, 2D and 3D respectively, (c) both stretched exponential attenuation and Mittag-Leffler function attenuation, the values for ($D_{f_1\parallel}$, $D_{f_1\perp}$) are ($D_{f_i 0}$, $10^6 D_{f_i 0}$), ($10^6 D_{f_i 0}$, $D_{f_i 0}$), ($D_{f_i 0}$, $D_{f_i 0}$) and ($D_{f_i 0}$, $2 D_{f_i 0}$), where $D_{f_i 0}$ represents $D_{f_1 0}$ or $D_{f_2 0}$. Other parameters used are $\alpha = 0.7, \beta =1.5$ and $D_{f_2 0} = D_{f_1 0}\Gamma(1+\alpha) = 6.57 \times 10^{-7}$ m$^\beta$/s$^\alpha$.

**Figure 1**

(a)
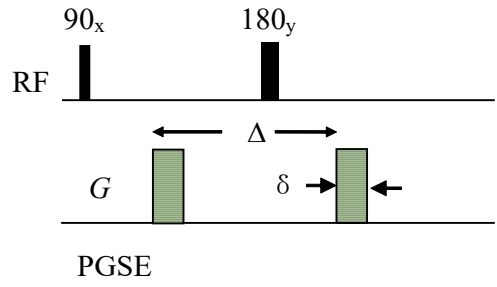
PGSE

(b)
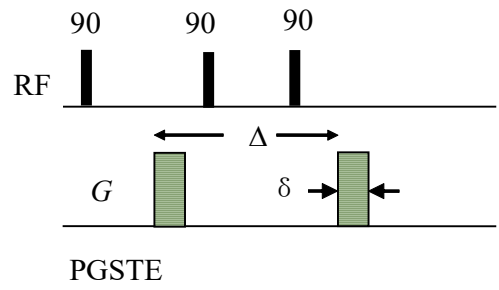
PGSTE

**Figure 2.**

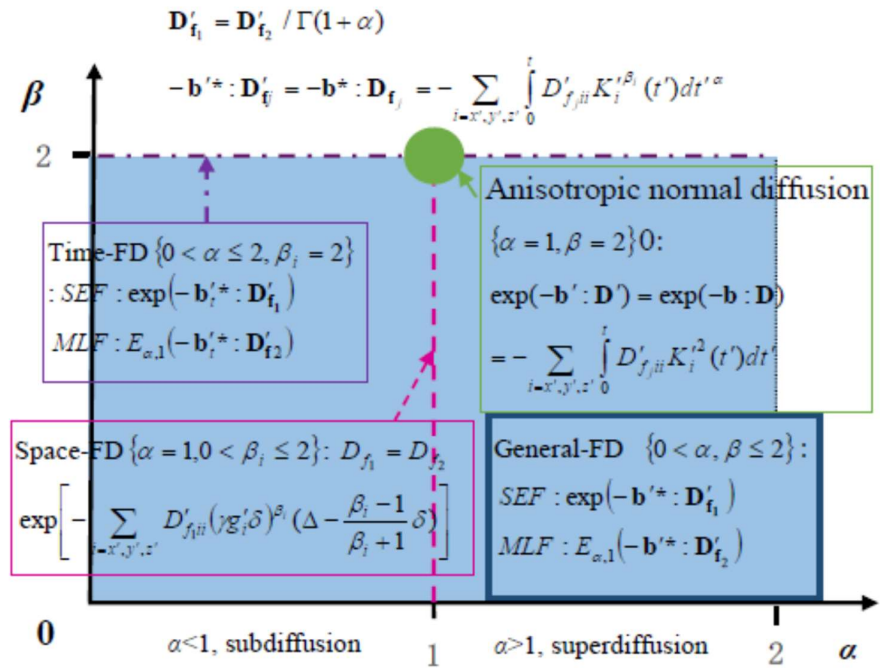

**Figure 3 (a)**

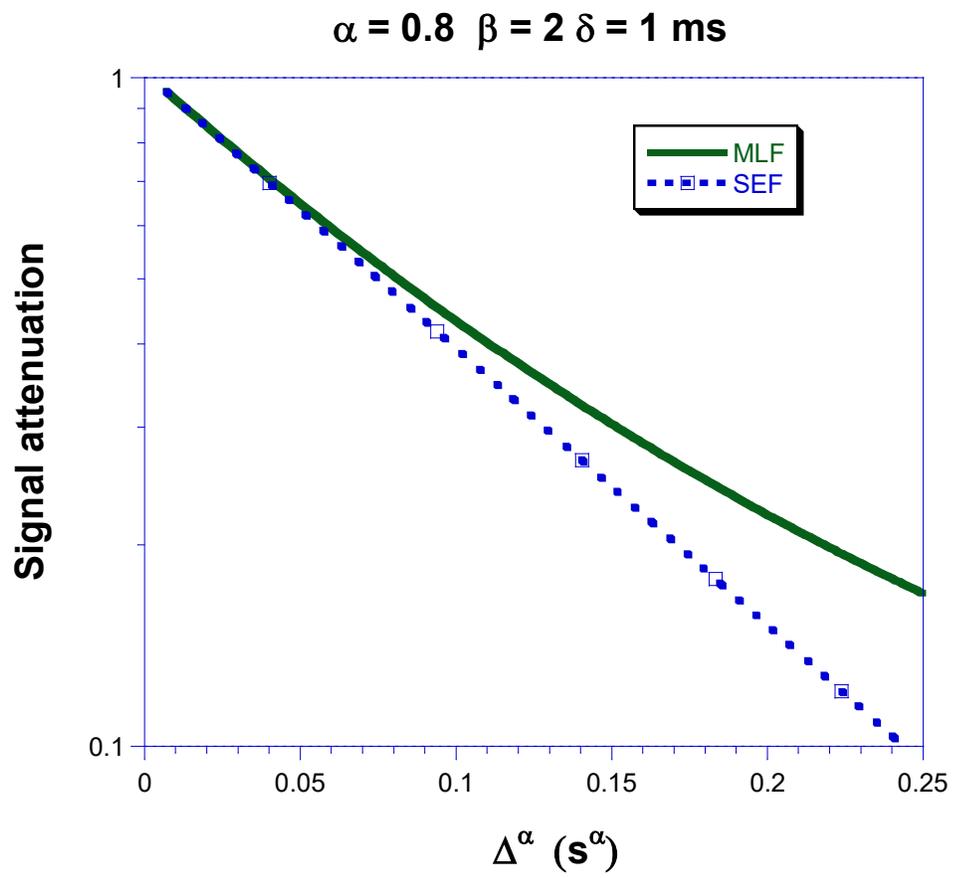

**Figure 3 (b)**

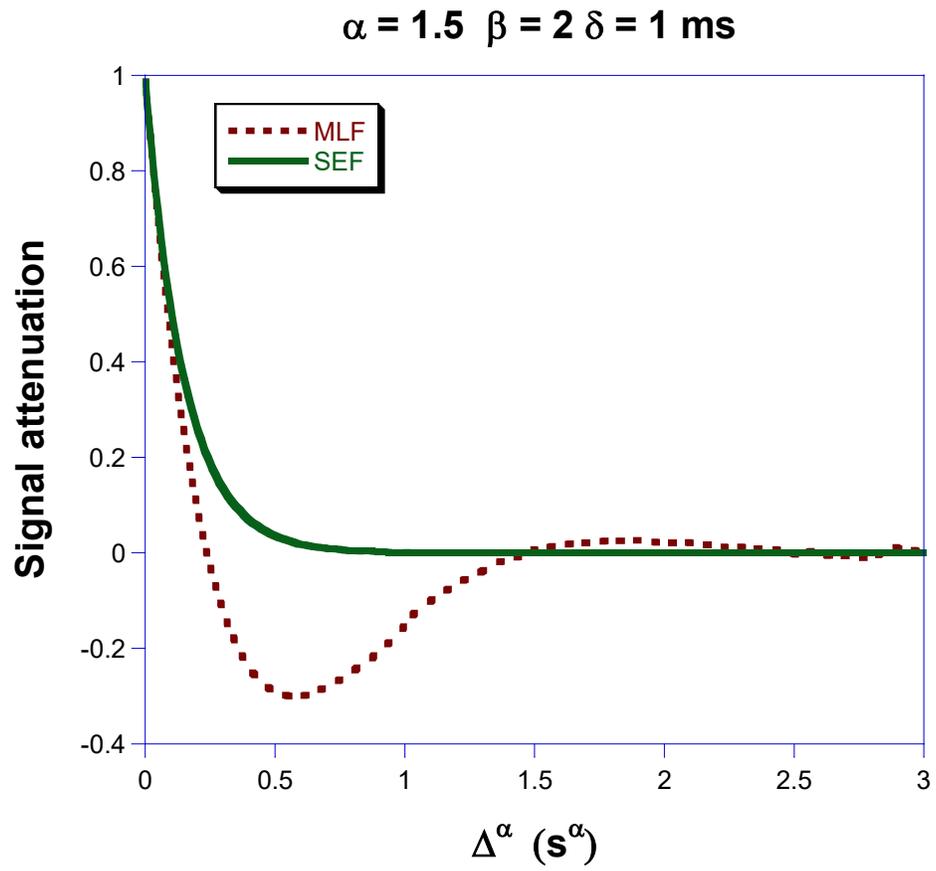

t^1.5

**Figure 3 (c)**

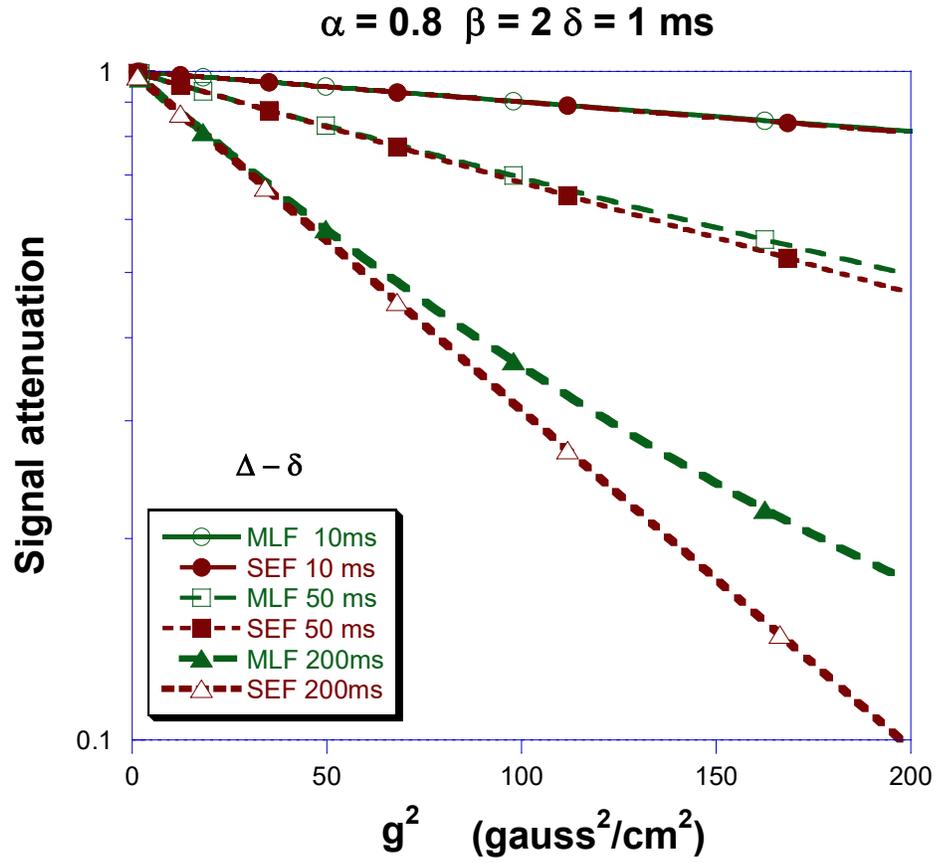

ani_vg_fgpw_subsuper_112416_0.8_2._1.38e-9_1ms_200ms_1_1

**Figure 3 (d)**

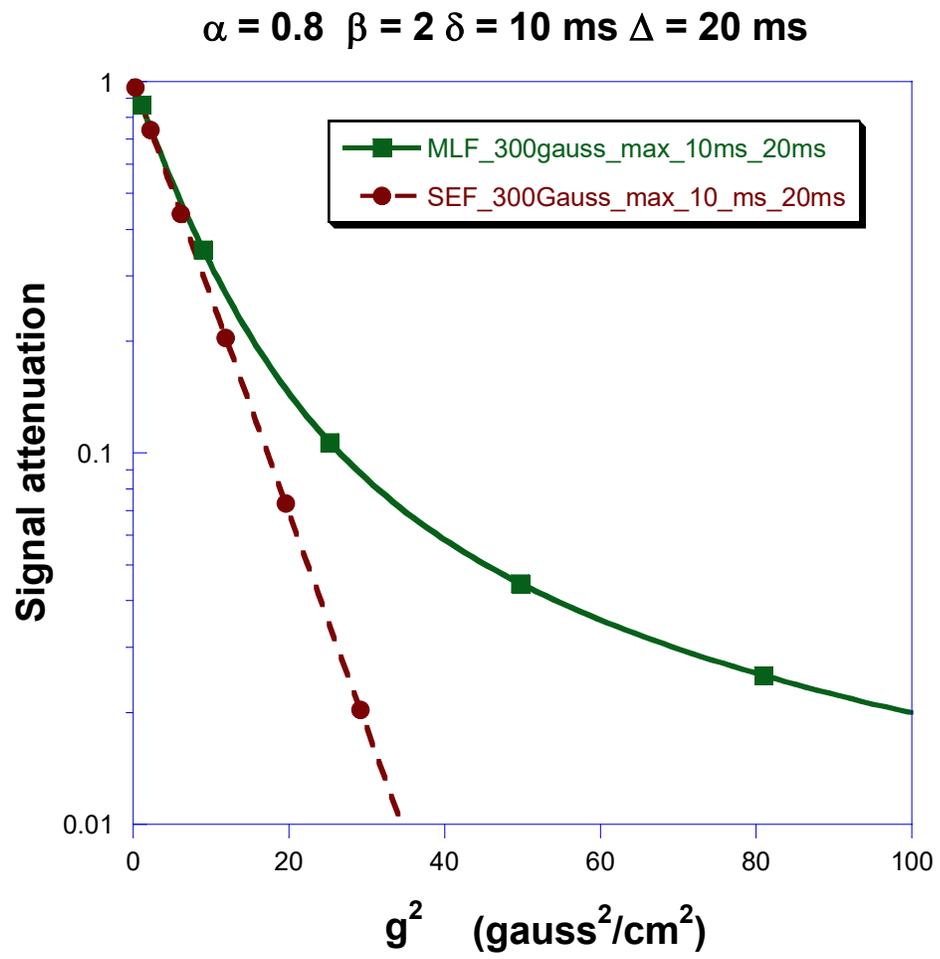

Guoxing Lin, PFG signal attenuation of anisotropic anomalous diffusion    24

**Figure 4 (a)**

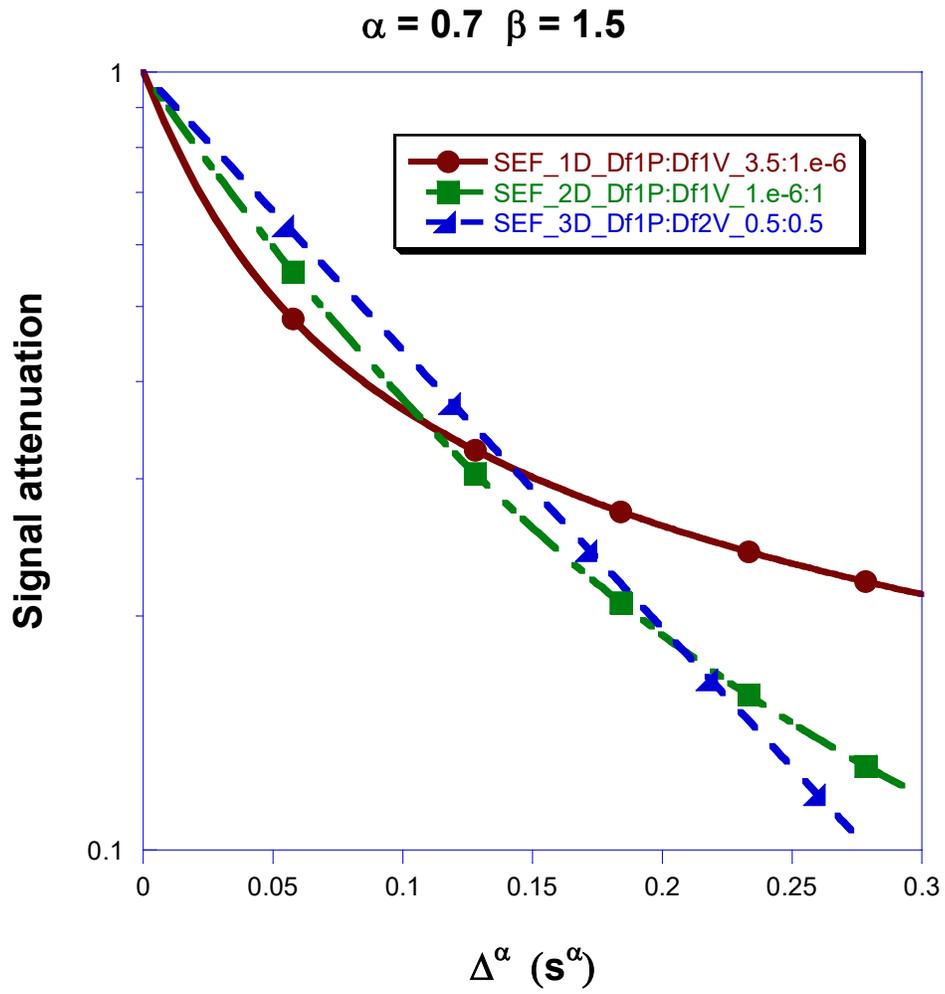

**Figure 4 (b)**

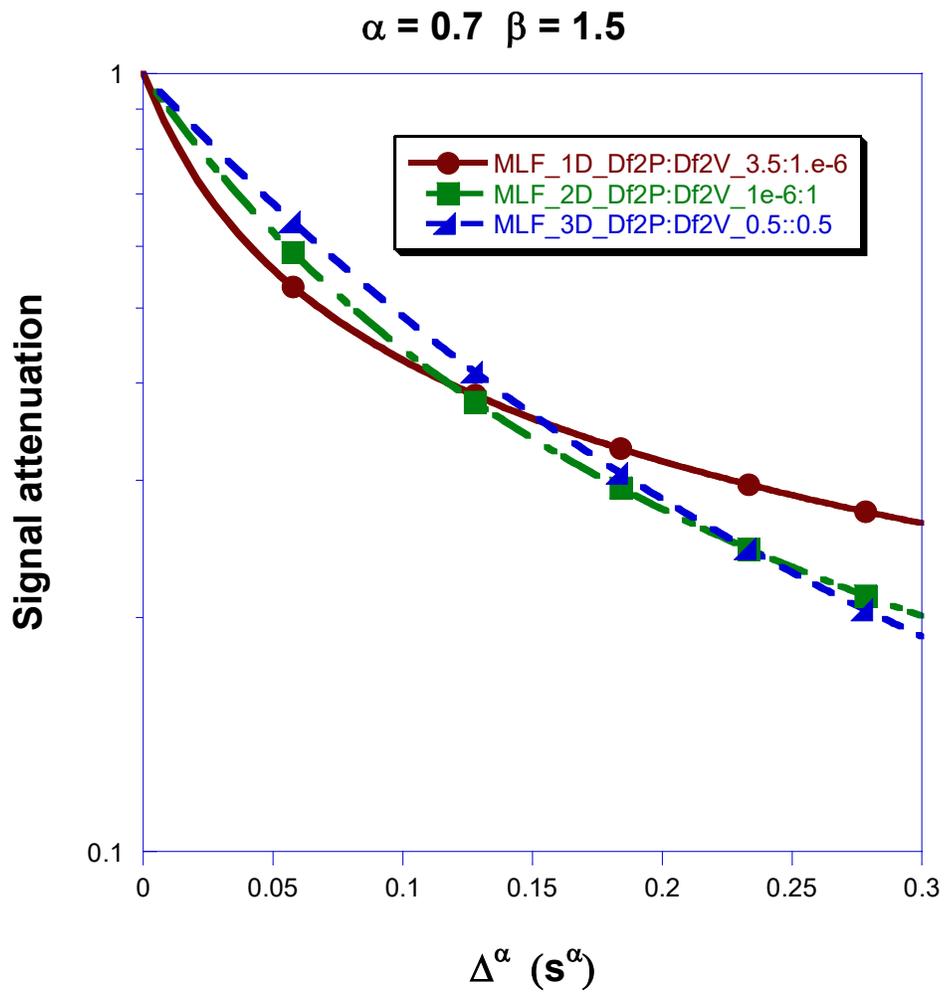

**Figure 4 (c)**

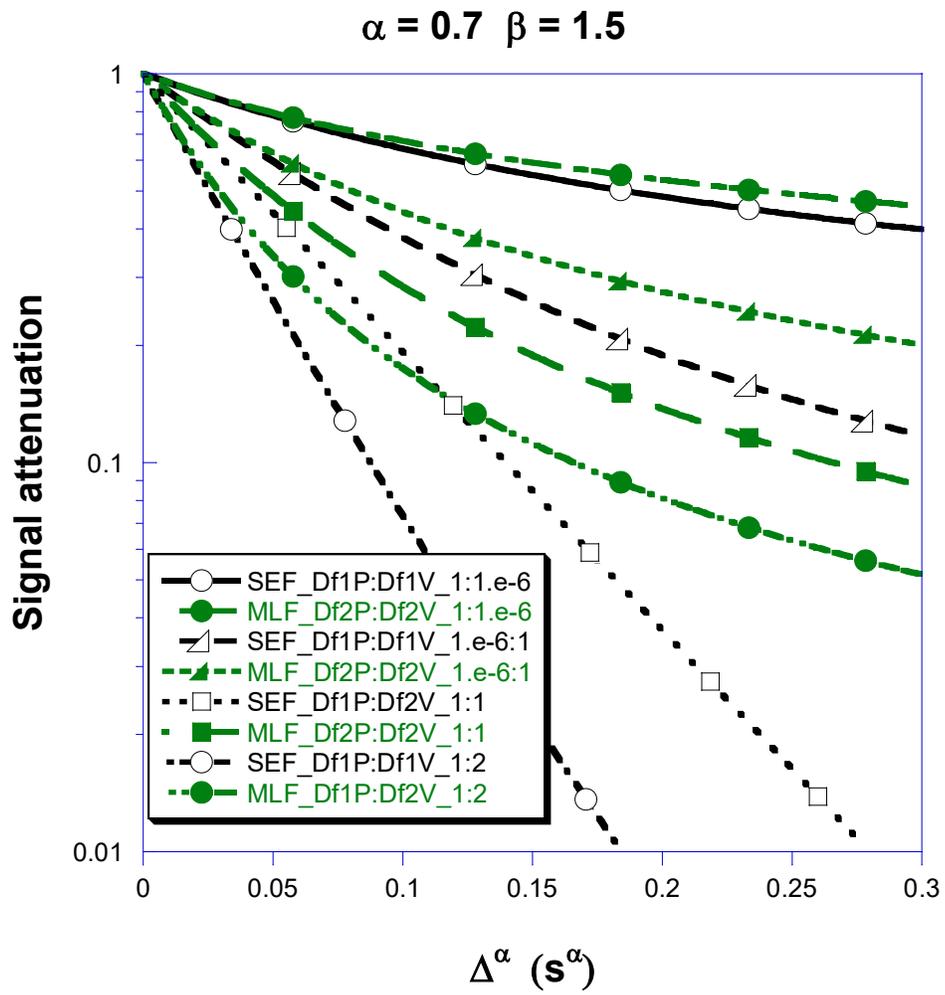